\documentclass{ipol}

\ipolSetTitle{Meta-analysis parameters computation: a Python approach to facilitate the crossing of experimental conditions}
\ipolSetAuthors{Flavien QUIJOUX\ipolAuthorMark{1}\ipolAuthorMark{2},
                Charles TRUONG\ipolAuthorMark{1},
                Ali\'enor VIENNE-JUMEAU\ipolAuthorMark{1},
                Laurent OUDRE\ipolAuthorMark{1}\ipolAuthorMark{3},
                François BERTIN-HUGAULT\ipolAuthorMark{2},
                Philippe ZAWIEJA\ipolAuthorMark{2},
                Marie LEFEVRE\ipolAuthorMark{2},
                Pierre-Paul VIDAL\ipolAuthorMark{4}\ipolAuthorMark{1},
                Damien RICARD\ipolAuthorMark{1}\ipolAuthorMark{5}\ipolAuthorMark{6}}
\ipolSetAffiliations{%
\ipolAuthorMark{1} Centre Borelli CNRS UMR 9010, Université Paris Descartes, Sorbonne Paris Cité, Service de Santé des Armées, 45 rue des Saints Pères, 75006 PARIS, France
                   (\texttt{f.quijoux@orpea.net})\\
\ipolAuthorMark{2} ORPEA Group, 12 rue Jean Jaurès, CS 10032, 92813 Puteaux Cedex, France\\
\ipolAuthorMark{3} LT2I, Université Paris 13, 99 avenue Jean-Baptiste Clément, 93430 Villetaneuse, France\\
\ipolAuthorMark{4}Institute of Information and Control, Hangzhou Dianzi University, Zhejiang, 310018, China\\
\ipolAuthorMark{5}Service de Neurologie de l’Hôpital d’Instruction des Armées de Percy, Service de Santé des Armées, 101 avenue Henri Barbusse, 92140 Clamart, France\\
\ipolAuthorMark{6}Ecole du Val-de-Grâce, Ecole de Santé des Armées, 1 Place Alphonse Laveran, 75005 Paris, France
}

\usepackage{amsmath}
\usepackage{amssymb}
\usepackage{dirtree}  

\begin{document}

\begin{ipolAbstract}
Meta-analysis is a data aggregation method that establishes an overall and objective level of evidence based on the results of several studies. It is necessary to maintain a high level of homogeneity in the aggregation of data collected from a systematic literature review. However, the current tools do not allow a cross-referencing of the experimental conditions that could explain the heterogeneity observed between studies. This article aims at proposing a Python programming code containing several functions allowing the analysis and rapid visualization of data from many studies, while allowing the possibility of cross-checking the results by experimental condition.
\end{ipolAbstract}

\ipolKeywords{meta-analysis, forest plot, sensitivity analysis}

\newpage
\section{Introduction}

Meta-analysis is a popular statistical procedure used to combine the results of several clinical studies that address the same research question~\cite{del_re_practical_2015}.
The objective of this method is to mitigate the bias associated with a particular selection of participants~\cite{israel_guide_2011} in a single study and increase the statistical significance of the conclusions~\cite{glass_primary_1976}.
Systematic reviews with meta-analysis are often seen as the highest level of evidence~\cite{10.1001/jama.1992.03490170092032}.
The main output is a quantitative measure of the effect of a treatment or medical condition, the so-called the effect size.
The statistical significance of this effect size is computed through a rigorous statistical approach which allows researchers to conclude on the presence or absence of a certain effect.
In addition, a sensitivity analysis is often be included in meta-analyses to compare the impact of the different experimental conditions~\cite{cook_methodologic_1995}.
Indeed, collecting studies that scrupulously share the same recording conditions can be practically challenging, leading to biased or heterogeneous selections.
On the other hand, thanks to the sensitivity analysis, researchers can identify spurious results (comparatively to other similar studies)~\cite{russo_how_2007, cook_methodologic_1995}, and therefore take action to mitigate the common effect.

\paragraph{Related work.}
To facilitate data aggregation and allow researchers to easily perform meta-analyses, the Cochrane Community has developed a software called Review Manager (RevMan)~\cite{cochrane2008review} which facilitates the writing of systematic reviews and the comparison of results from scientific articles.
RevMan has a graphical interface and is particularly well disseminated in the scientific community because this software integrates many functionalities, including the writing of the review itself, the production of graphs, and the evaluation of bias~\cite{the_cochrane_collaboration_revman_2014}.
However, Cohrane's software has two main drawbacks.
First, the exact formulas to compute the graphs are not easily found in the associated handbook.
Second, when dealing with several subgroups in the clinical studies, researchers have to manually enter several time the same data, leading to potential copy errors.
The amount of manual data duplication quickly grows as the number of considered subgroups increases.
As an example, the systematic review in~\cite{quijoux_center_2020}, considers 29 studies, 26 different variables of interest, and 8 experimental parameter values.
This yields a table of 241 lines, where each line corresponds to a study, a variable and a combination of experimental conditions.
We measured that in total, 297 different meta-analyses could be performed, by selecting all subsets of experimental settings.
To compute all analyses in RevMan, we would need to manually enter each line about 6 times in the software, a procedure which can be error prone.
This difficulty of combining results due to variations in experimental conditions has been previously highlighted in systematic reviews~\cite{piirtola_force_2006,quijoux_center_2019,ruhe_center_2011}. For instance, in the field of postural control analysis, the number of extracted parameters can exceed one hundred~\cite{comber_postural_2018}, making the subgroups analysis for each outcome and each condition of recording with RevMan unpractical.

\paragraph{Contributions.} 
We aim at providing an easy-to-use tool for conducting a meta-analysis even when the number of conditions and experimental subgroups is large. Thanks to this work, quick and easy selection of data according to the experimental conditions in which they were recorded can be performed in one operation, without manual data duplication.
In addition, all calculations of the overall effect size, confidence interval, weights and model selection are carefully explained.

\subsection{Outline}
Section~\ref{sec:methodo} describes the general objective of meta-analysis and provides an illustrative example.
In the section 3, the exact steps to computed the overall effect size, the associated confidence interval, the forest and funnel plots are formally explained.
The last section presents the software that goes with this article, especially the computational requirements, and the input and output formats.
\section{Methodology of a meta-analysis}\label{sec:methodo}

This section presents a general description of a meta-analysis and a simplified example to illustrate notions that will be used later as well as further motivate the need for tools which can deal with several experimental settings.

\subsection{General principle}

The objective of the meta-analysis is to determine if an outcome is significantly different between two groups (typically, a experimental or intervention group and a control group).
The outcome is measured by one or several variables of interest, whose values (means and standard deviations) are measured on each group and compared.
Depending on whether the difference between both groups is significantly away from zero, one can then conclude on the presence or absence of influence between the outcome and the group membership.
By combining results from individual clinical studies, a systematic review can summarize and provide more robust measures of influence.
The first step to a meta-analysis is to gather similar clinical studies that try to answer the same research question and extract the relevant information.
Roughly, an article is included in a systematic review if it measures one of the variables of interest, in similar experimental settings~ \cite{israel_guide_2011}.
The relevant information consist in the number of participants, the means and standard deviations of the considered variables, as well as contextual information about the clinical protocol.
After this data collection process, it is possible to quantify the influence of a group to the considered outcome, by computing the so-called effect size which is, for a given variable of interest, proportional to the standardized mean difference.
This quantity is computed, along with confidence intervals, for each individual study as well as for the considered pool of articles (the overall effect size).
Since the results extracted from the selected studies contain some heterogeneity (random or not), extracted values are weighted according to the size of the associated cohorts and the variability of the measures.
All computations (effect size and confidence interval) are often summarized in a visual representation, called a forest plot~\cite{10.1093/ije/dyp370}, which allows to visualize the relative contribution of each article to the overall effect size, and to quickly assess if a combined studies confirm the influence of a group to the outcome.
In addition, a sensitivity analysis is also performed, to detect selection bias, which can occur because of a higher probability to publish significant results in the scientific literature.
To that end, another visual representation, called a funnel plot is drawn.
It consists of a scatter diagram where each study is represented according to their intra-study effect size and variance.

\subsection{Illustrative example: the study of quiet stance balance and fall risk}
(A complete version of this simplified meta-analysis protocol and results can be found in~\cite{quijoux_center_2020}.)
\\
In order to assess the risk of fall in the elderly population (which is one of the main causes of deadly injuries in this population~\cite{world_health_organization_who_2008}), the different postural strategies of senior individuals are analyzed~\cite{audiffren_non_2016,bargiotas_importance_2018,bloch_estimation_2013,muir_quantifying_2010}.
To that end, medical researchers often focus on the static balance of patients, measured by force platforms~\cite{quijoux_center_2019}.
Force platforms record the displacement of the center of pressure (COP), i.e.\ the resultant of the weight distribution between the two legs, and from that recording, several quantities (velocity, duration, covered surface, etc.) are computed.
Generally, participants are asked to remain stable on the platform; however, a great variability in the experimental conditions of the recording can be observed, leading to possibly many different meta-analyses, depending on which experimental setting is preferred.
Here, we focus on the COP mean velocity, one of the most common features, especially in the anteroposterior (AP) direction, and two types of experimental conditions: eyes open or closed (the participant had his or her eyes open or closed during the recording) and retrospective or prospective fall recordings (a retrospective history of past falls occurred before the recording or a prospective follow-up of the falls occurred after the recording).
The objective of a meta-analysis is to determine the relationship between the variable of interest (here, the AP mean velocity) and the risk of fall (here, the population is simply divided into fallers and non-fallers).
Seven clinical studies are included in this meta-analysis and Table~\ref{tab:example_data} summarizes the information extracted from the associated articles, namely the number of participants, the mean and standard deviations of the group of fallers and of the group of non-fallers, as well as the experimental conditions.
\begin{table}[t!]
    \centering
    \begin{tabular}{l|l|c|c|c|c|c|c}
        Study & Condition $C$ & $n_1^{(k)}(\theta)$ & $\mu_1^{(k)}(\theta)$ & $\sigma_1^{(k)}(\theta) $ & $n_2^{(k)}(\theta)$ & $\mu_2^{(k)}(\theta)$ & $ \sigma_2^{(k)}(\theta)$ \\ \hline
        Howcroft, 2015 & EO $\times$ Retro  & 24    & 7.34  & 2.47  & 76    & 7.65 & 1.84 \\
        Howcroft, 2015 & EC $\times$ Retro  & 24    & 17.34 & 16.03 & 76    & 15.86 & 6.74 \\
        Howcroft, 2017 & EC $\times$ Pro    & 42    & 17.76 & 13.4  & 47    & 15.11 & 5.59 \\
        Howcroft, 2017 & EO $\times$ Pro    & 42    & 7.75  & 2.15  & 47    & 7.53 & 1.93 \\
        König, 2014 & EC $\times$ Retro     & 42    & 0.15  & 1.48  & 42    & -0.12 & 0.12 \\
        Kwok, 2015 & EO $\times$ Retro      & 18    & 1.27  & 0.45  & 55    & 1.02 & 0.26 \\
        Maki, 1994 & EC $\times$ Pro        & 59    & 17.9  & 15.6  & 37    & 11.9 & 4.79 \\
        Maki, 1994 & EO $\times$ Pro        & 59    & 13    & 13.7  & 37    & 8.4 & 3.51 \\
        Maranesi, 2016 & EC $\times$ Retro  & 63    & 16.23 & 11.27 & 67    & 14.5 & 9.1 \\
        Maranesi, 2016 & EO $\times$ Retro  & 63    & 11    & 6.89  & 67    & 10 & 6.2 \\
        Pajala, 2008 & EC $\times$ Pro      & 189   & 12.46 & 5.09  & 230   & 12.5 & 6.8 \\
        Pajala, 2008 & EO $\times$ Pro      & 189   & 8.34  & 2.81  & 230   & 7.8 & 2.6 
    \end{tabular}
    \caption{Input example for the proposed meta-analysis procedure.
    Here a single variable of interest is considered (``Anteroposterior mean velocity'').
    ``EO'' and ``EC'' respectively denote ``Eyes Open'' and ``Eyes Closed''.
    ``Retro'' corresponds to a retrospective fall history (before the recording) and ``Pro'' corresponds to a prospective follow-up of the falls (after the recording).
    The group $1$ is the group of elderly fallers; the group $2$ is the group of elderly non-fallers.
    The group sample sizes $n_1^{(k)}(\theta)$ and $n_2^{(k)}(\theta)$, the means $\mu_1^{(k)}(\theta)$ and $\mu_2^{(k)}(\theta)$, the standard deviations $\sigma_1^{(k)}(\theta)$ and $\sigma_2^{(k)}(\theta)$ are extracted from the associated articles.
    (See Section~\ref{sec:notations} for the notations.)
    }
    \label{tab:example_data}
\end{table}%
%
%

\begin{figure}[t!]
    \centering
    \begin{minipage}[t]{\linewidth}
    \centerline{\includegraphics[width=0.72\linewidth]{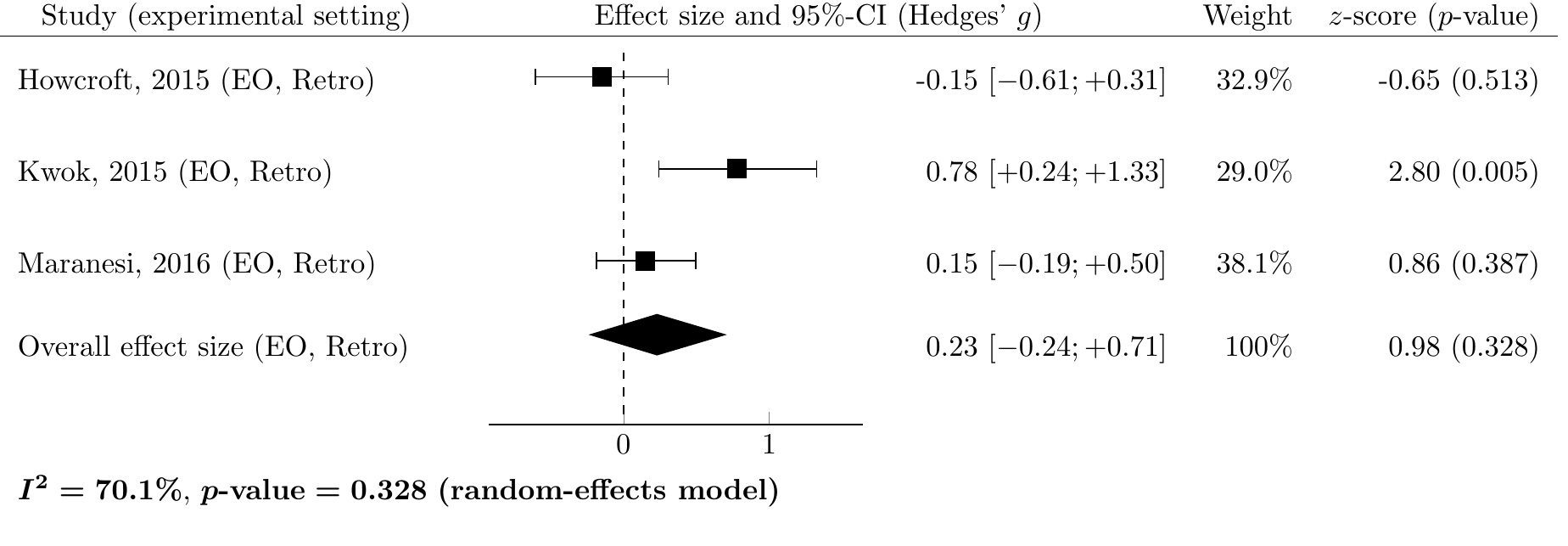}}
    \centerline{\footnotesize (a) Experimental setting: eyes open (``EO'') and retrospective fall recording (``Retro'').}
    \end{minipage}
    \vskip1em
    \begin{minipage}[t]{\linewidth}
    \centerline{\includegraphics[width=0.72\linewidth]{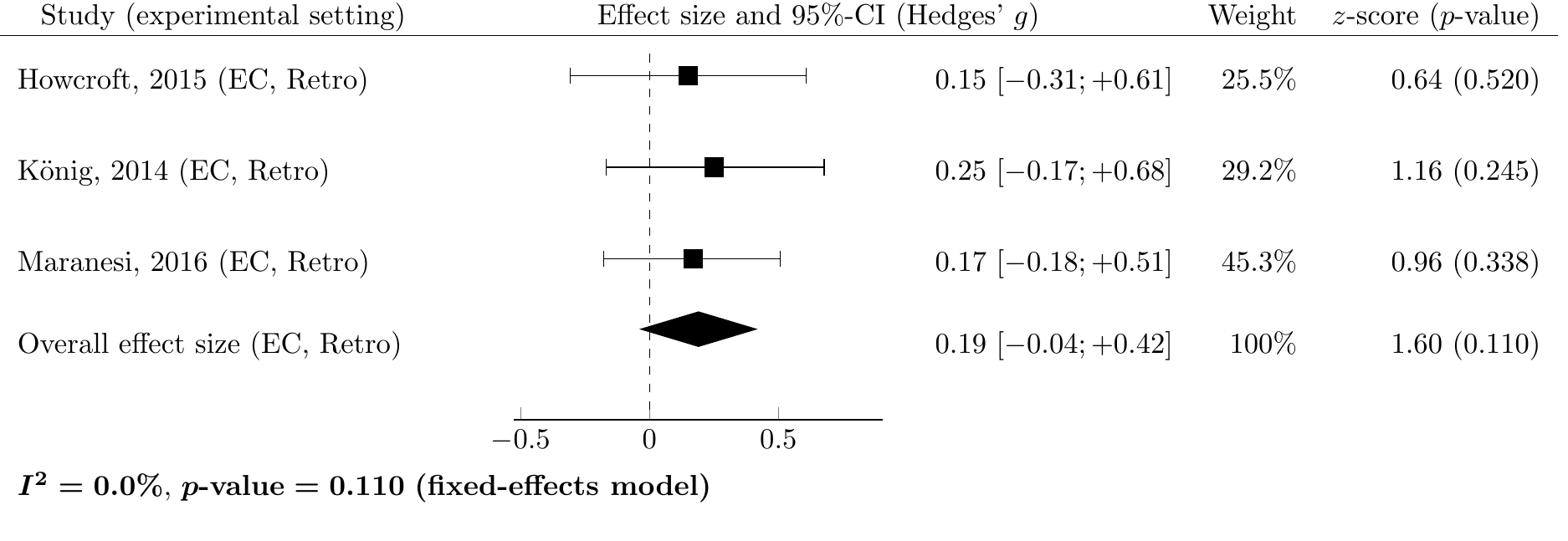}}
    \centerline{\footnotesize (b) Experimental setting: eyes closed (``EC'') and retrospective fall recording (``Retro'').}
    \end{minipage}
    \vskip1em
    \begin{minipage}[t]{\linewidth}
    \centerline{\includegraphics[width=0.72\linewidth]{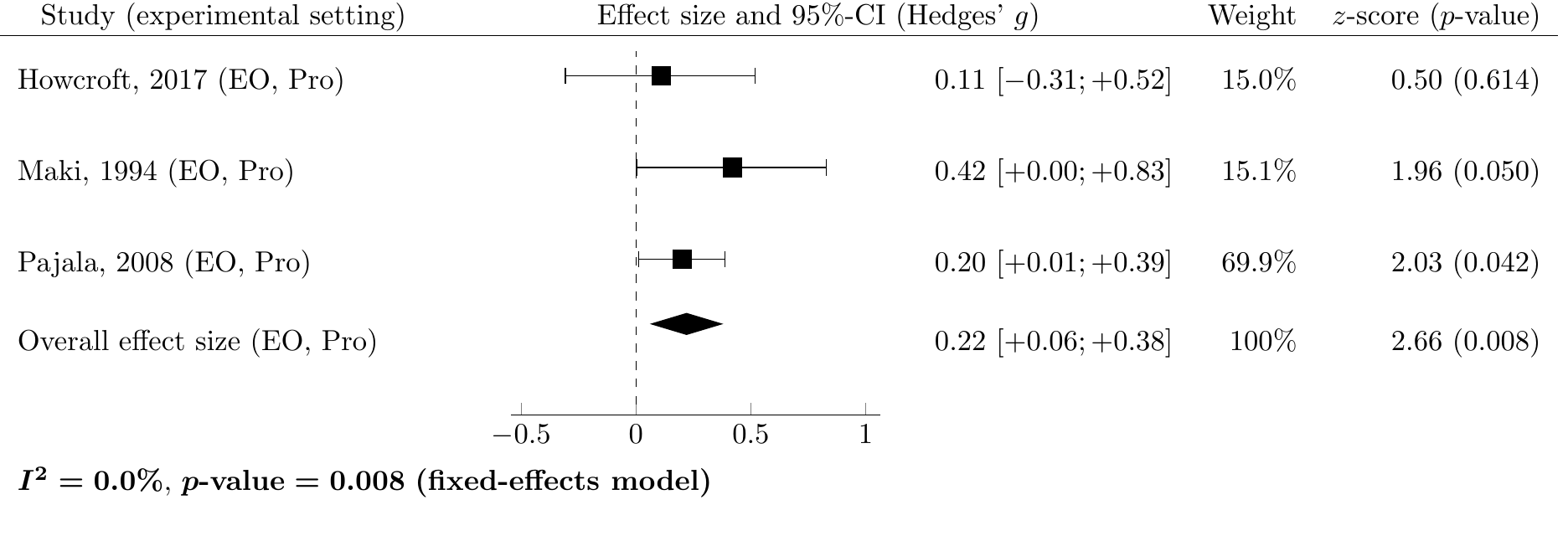}}
    \centerline{\footnotesize (c) Experimental setting: eyes open (``EO'') and prospective fall recording (``Pro'').}
    \end{minipage}
    \vskip1em
    \begin{minipage}[t]{\linewidth}
    \centerline{\includegraphics[width=0.72\linewidth]{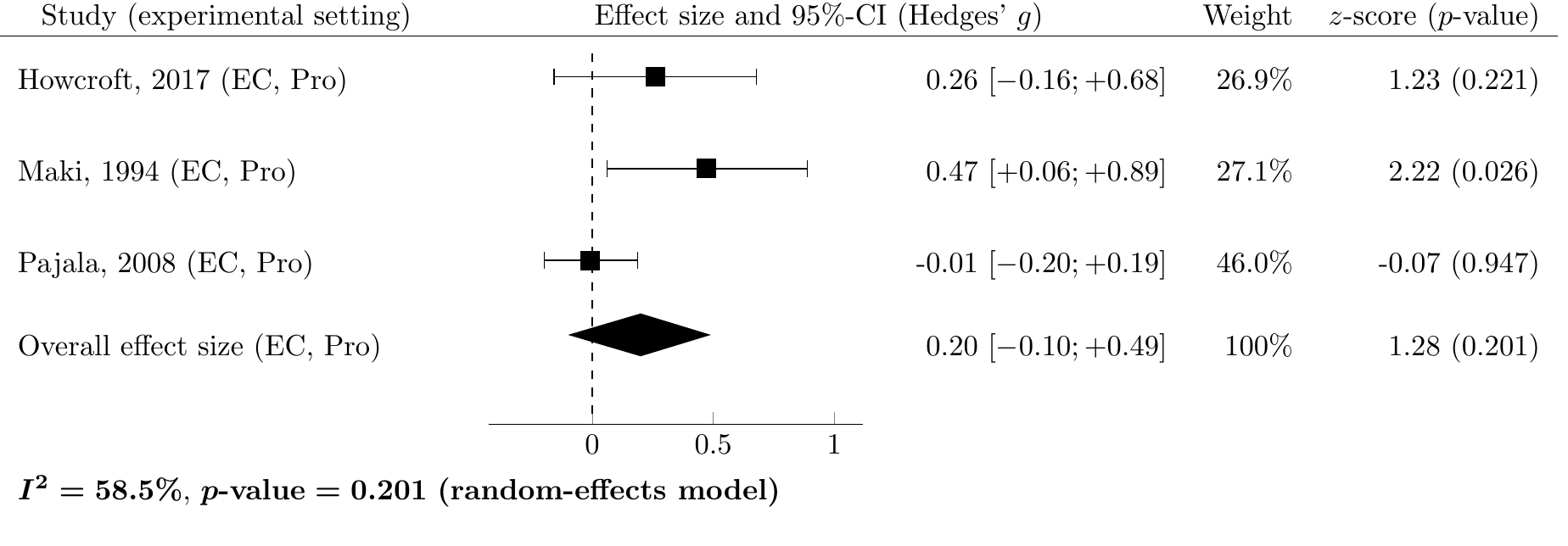}}
    \centerline{\footnotesize (d) Experimental setting: eyes closed (``EC'') and prospective fall recording (``Pro'').}
    \end{minipage}
    \caption{Forest plot for the input data of Table~\ref{tab:example_data}.
    The variable of interest is ``AP mean velocity''
    The effect sizes and associated $95\%$ confidence intervals are computed with our procedure for two sets experimental conditions (eyes open or closed, retrospective or prospective fall recording).
    Note that, for brevity, only 4 out of the 8 possible combinations of experimental settings are shown here.}
    \label{fig:example-forest-plots}
\end{figure}

\noindent
From the 12 measures in Table~\ref{tab:example_data}, 8 distinct meta-analyses can be performed, one for each combination of conditions\footnote{In detail: EO, EC, Pro, Retro, EC$\times$Pro, EC$\times$Retro, EO$\times$Pro, EO$\times$Retro.}.
With RevMan, this would have resulted in duplicating manually those 12 lines in order to compute all of those analyses because it is not designed to cope with multiple combinations of conditions.
The meta-analysis tool that we propose is able to compute all conditions in one pass.
In total, for this simple example, 36 lines would have been entered into RevMan while only 12 are needed with our method.
Part of the usual output of meta-analyses, namely the forest plots, is shown on Figure~\ref{fig:example-forest-plots}.
Those forest plots illustrate the sensitivity of the AP mean velocity to the different conditions.
The effect size and its $95\%$ confidence interval is provided for each study, along with the overall effect size (diamond shape).
The vertical dashed line represents the absence of effect, and intuitively, if a confidence interval of an individual source crosses this line, it means that there is no difference between the two groups for this study.
Conversely, if the confidence interval of the overall effect size crosses the null-effect line, it indicates that the meta-analysis does not demonstrate a significant effect for the variable of interest.

\section{Computing the effect sizes}

This section presents the computation of the individual effect size, calculated for each study, and the overall effect size, calculated over several studies sharing the same experimental setting.

\subsection{Notations}\label{sec:notations}
A subgroup of studies is defined by a variable of interest $X$ and a set of experimental conditions $C$.
All articles that evaluate this variable $X$ under the conditions $C$ form the subgroup.
For ease of notation, the pair $[X, C]$ is denoted by $\theta=[X, C]$ in the remaining of the article.
The number of articles in the subgroup defined by $\theta$ is denoted $K(\theta)$ ($K(\theta)>1$).
For each study $k$ ($k=1,\dots,K(\theta)$), a user needs to extract the following quantities:
\begin{itemize}
    \item $n_{1}^{(k)}(\theta)$, $\mu_{1}^{(k)}(\theta)\in\mathbb{R}$, $\sigma_{1}^{(k)}(\theta)\in\mathbb{R}_+$, respectively the number of subjects, the empirical mean and the (unbiased) standard deviation of $X$, for the first group (typically the study group),
    \item $n_{2}^{(k)}(\theta)$, $\mu_{2}^{(k)}(\theta)\in\mathbb{R}$, $\sigma_{2}^{(k)}(\theta)\in\mathbb{R}_+$, respectively the number of subjects, the empirical mean and the (unbiased) standard deviation of $X$, for the second group (typically the control group).
\end{itemize}

\subsection{Individual effect size}
Roughly speaking, for a given study $k$, the individual effect size quantitatively measures how the variable $X$ of interest varies between the two considered groups of subjects.
For instance, this measure could determine if access to a certain treatment influences a certain measure of health status~\cite{littell_systematic_2009}.
Two definitions of this quantity coexist: Cohen's, denoted $\delta_d$ \cite{cohen_statistical_1988}, and Hedges', denoted $\delta_g$ \cite{hedges_distribution_1981}.
(In the literature, those two quantities are commonly referred to as Cohen's $d$ and Hedges' $g$.)
The former is simply equal to the standardized mean difference
\begin{equation}
\delta_d^{(k)}(\theta)=\frac{\mu_{1}^{(k)}(\theta) -\mu_{2}^{(k)}(\theta)}{S^{(k)}(\theta)}.
\label{eq:cohen-es}
\end{equation}
where $\mu_{1}^{(k)}(\theta)$, $\mu_{2}^{(k)}(\theta)$ are extracted from the study $k$ and $S^{(k)}(\theta)$ is the weighted standard deviation~\cite{hedges2014statistical} defined by
\begin{equation}
S^{(k)}(\theta) =\sqrt{\frac{\left(n_{1}^{(k)}(\theta)-1\right) [\sigma_1^{(k)}(\theta)]^2+\left(n_2^{(k)}(\theta)-1\right) [\sigma_2^{(k)}(\theta)]^2}{n_1^{(k)}(\theta)+n_2^{(k)}(\theta)-2}}
\label{eq:weightedpooledsd}
\end{equation}
with $\sigma_1^{(k)}$, $\sigma_2^{(k)}$, $n_1^{(k)}$, and $n_2^{(k)}$ are extracted from study $k$.
%
%
This statistic is known to be upwardly biased with small samples~\cite{faraone_interpreting_nodate}.
As a results, Hedges' $\delta_g$ has been introduced to better estimate effect size, even for studies with only few samples~\cite{grissom_effect_2005}.
This measures is defined by
\begin{equation}
\delta_g^{(k)}(\theta)=\left(1-\frac{3}{4[n_1^{(k)}(\theta)+n_2^{(k)}(\theta)]-9}\right) \times\frac{\mu_1^{(k)}(\theta)-\mu_2^{(k)}(\theta)}{S^{(k)}(\theta)}.
\label{eq:hedges-es}
\end{equation}
In the literature, both statistics, Cohen's $\delta^{(k)}_d$~\eqref{eq:cohen-es} or Hedges' $\delta^{(k)}_g$~\eqref{eq:hedges-es}, can be used to estimate the individual effect size and it is often left to the user to choose between one or the other.
In the remaining of the article, $\delta^{(k)}$ will denote either $\delta^{(k)}_d$ or $\delta^{(k)}_g$.

\subsection{Overall effect size}
\label{subsec:overall-es}
In a nutshell, the overall effect size across studies, denoted $\mu(\theta)$, is a weighted average of the individual effect sizes:
\begin{equation}
\mu(\theta)=\frac{\sum\limits_{k=1}^{K(\theta)} w^{(k)}(\theta) \times \delta^{(k)}(\theta)}{\sum\limits_{k=1}^{K(\theta)}w^{(k)}(\theta)}
\label{eq:overall-es}
\end{equation}
where $\delta^{(k)}$ can either be $\delta^{(k)}_d$~\eqref{eq:cohen-es} or $\delta^{(k)}_g$~\eqref{eq:hedges-es}, and the weights $w^{(k)}$ are introduced in the following sections. Roughly speaking, depending on the heterogeneity between studies (the inter-study variability), one can choose between a fixed-effect model, in which case $w^{(k)}=w^{(k)}_{FE}$ (Section~\ref{sec:fe-model}), or a random-effect model, in which case $w^{(k)}=w^{(k)}_{RE}$ (Section~\ref{sec:re-model}).
Both are described below, along with a criterion to choose between them.

\subsubsection{Heterogeneity between studies: fixed-effects model}\label{sec:fe-model}
The fixed-effects model assumes that the set of considered studies are homogeneous, meaning that the differences between the extracted values $\mu_i^{(k)}$ and $\sigma_i^{(k)}$ ($i\in\{1, 2\}$) for varying $k$ are only the result of random noise~\cite{del_re_practical_2015, Rileyd549}.
In this context, where no heterogeneity between articles is considered and all studies estimate the exact same effect, the weights of the fixed-effects model, denoted by $w_{FE}^{(k)}$, are defined by
\begin{equation}
w_{FE}^{(k)}(\theta)=\frac{1}{[\sigma_{\text{intra}}^{(k)}(\theta)]^2}
\label{eq:weight-fe}
\end{equation} 
where $\sigma_{\text{intra}}^{(k)}$ is the intra-study variance:
\begin{equation}
\sigma_{\text{intra}}^{(k)}(\theta)= \left(\frac{n_1^{(k)}(\theta)+n_2^{(k)}(\theta)}{n_1^{(k)}(\theta) \times n_2^{(k)}(\theta)} + \frac{[\delta^{(k)}(\theta)]^2}{2[n_1^{(k)}(\theta)+n_2^{(k)}(\theta)]}\right)^{1/2}
\label{eq:intra-var}
\end{equation}
where $\delta^{(k)}$ can either be $\delta_d^{(k)}$~\eqref{eq:cohen-es} or $\delta_g^{(k)}$~\eqref{eq:hedges-es}.
According to this model, studies with a large number of samples ($n_i^{(k)}$, $i\in\{1, 2\}$) have a large weight $w_{FE}^{(k)}$~\eqref{eq:weight-fe} and thus carry more information.
Equation~\ref{eq:overall-es} can now be rewritten and the overall effect size for a fixed-effects model, denoted $\mu_{FE}(\theta)$, is 
\begin{equation}
\mu_{FE}(\theta)=\frac{\sum\limits_{k=1}^{K(\theta)} w_{FE}^{(k)}(\theta) \times \delta^{(k)}(\theta)}{\sum\limits_{k=1}^{K(\theta)}w_{FE}^{(k)}(\theta)}.
\label{eq:overall-es-fe}
\end{equation}

\subsubsection{Heterogeneity between studies: random-effects model}\label{sec:re-model}

Whenever there is heterogeneity between studies, for instance because of differences in how measurements are taken or in how the variable of interest is computed~\cite{ades_interpretation_2005}, the fixed-effects model no longer applies: one can then resort to the random-effects model~\cite{dersimonian_meta-analysis_1986}.
Before defining the weights of this model, two quantities are now introduced: the heterogeneity measure $Q(\theta)$ and the inter-study variance $\tau(\theta)$.
First, the heterogeneity measure $Q(\theta)$ is derived from the fixed-effects model:
\begin{equation}\label{eq:hetero-measure}
\mathcal{}Q(\theta)=\sum\limits_{k=1}^{K(\theta)} w_{FE}^{(k)}(\theta) \times [\delta^{(k)}(\theta)-\mu_{FE}(\theta)]^2.
\end{equation}
Second, the inter-study variance is as follows:
\begin{equation}
\mathcal{}\tau(\theta)^2=\frac{Q(\theta)-(K(\theta)-1)}{\xi(\theta)}
\label{eq:tau}
\end{equation}
where the coefficient $\xi$ is computed from the following equation: 
\begin{equation}
\xi(\theta)=\sum\limits_{k=1}^{K(\theta)}w_{FE}^{(k)}(\theta) - \frac{\sum\limits_{k=1}^{K(\theta)} [w_{FE}^{(k)}(\theta)]^2}{\sum\limits_{k=1}^{K(\theta)}w_{FE}^{(k)}(\theta)}.
\label{eq:xi}
\end{equation}
(Note that thus defined, $\tau^2$ is not guaranteed to be positive, therefore, in practice, negative $\tau^2$ are clipped to $0$ as in \cite{veroniki_methods_2016,higgins_commentary_2008}.)
The weights of the individual studies under the random-effects model are defined by
\begin{equation}
w_{RE}^{(k)}(\theta)=\frac{1}{[\sigma_{\text{intra}}^{(k)}(\theta)]^2 + [\tau(\theta)]^2}.
\label{eq:weight-re}
\end{equation}
Equation~\ref{eq:overall-es} can now be rewritten and the overall effect size for a random-effects model, denoted $\mu_{RE}(\theta)$, is 
\begin{equation}
\mu_{RE}(\theta)=\frac{\sum\limits_{k=1}^{K(\theta)} w_{RE}^{(k)}(\theta) \times \delta^{(k)}(\theta)}{\sum\limits_{k=1}^{K(\theta)}w_{RE}^{(k)}(\theta)}.
\label{eq:overall-es-re}
\end{equation}

\subsubsection{Choose between fixed-effects and random-effects}\label{sec:fe-or-re}
In order to choose between a fixed-effects model and a random-effects, the Cochrane~\cite{chandler_cochrane_2017} proposes a quantitative methodology based on the heterogeneity measure $Q(\theta)$ (Equation~\ref{eq:hetero-measure}), and more precisely, on a derived percentage, denoted $I^2(\theta)$~\cite{higgins_quantifying_2002,higgins_measuring_2003,del_re_practical_2015}:
\begin{equation}\label{eq:i-square}
I^2(\theta)=\frac{Q(\theta) - (K(\theta)-1)}{Q(\theta)} \times 100.
\end{equation}
According to the Cochrane~\cite{chandler_cochrane_2017}, when the value of $I^2$ is greater than 50\%, the inter-study variability is substantial and the random-effects model should be chosen.
Otherwise, the variability is considered to be moderate and the fixed-effects model should be preferred.
Note that thus defined, $I^2$ is not guaranteed to be positive, therefore, in practice, negative $I^2$ are clipped to $0$ as in \cite{higgins_measuring_2003,von_hippel_heterogeneity_2015}.

\subsection{Visualizations of a meta-analysis}
This section describes two visual tools commonly used by researchers to quickly assess the magnitude of the overall effect size and the contribution of each studies included in the meta-analysis, namely the forest plot and the funnel plot.

\subsubsection{Forest plot}
For a given $\theta$ (variable of interest and experimental condition), a forest plot displays the confidence intervals of the individual effect sizes, Cohen's $\delta^{(k)}_d$~\eqref{eq:cohen-es} or Hedges' $\delta^{(k)}_g$~\eqref{eq:hedges-es}, and of the overall effect size, $\mu_{FE}$~\eqref{eq:overall-es-fe} or $\mu_{RE}$~\eqref{eq:overall-es-re}. An example is shown on Figure~\ref{fig:example-forest-plots}. 
\\
To that end, the effect sizes are modeled by Gaussian random variables. For each study $k$, the confidence interval of the individual effect size, at the level $1-\alpha$, is
\begin{equation}\label{eq:interval-conf-delta}
[\delta^{(k)}(\theta) -\sigma^{(k)}_{\text{intra}}(\theta) q_{\alpha/2}, \delta^{(k)}(\theta) + \sigma^{(k)}_{\text{intra}}(\theta) q_{\alpha/2}]
\end{equation}
where $\delta^{(k)}$ stands for $\delta^{(k)}_d$~\eqref{eq:cohen-es} or $\delta^{(k)}_g$~\eqref{eq:hedges-es}, $\sigma^{(k)}_{\text{intra}}$ is the intra-study variance~\eqref{eq:intra-var} and $q_{\alpha/2}$ is the quantile function of a standard normal distribution.
Similarly, the confidence interval of the overall effect size, at the level $1-\alpha$, is 
\begin{equation}\label{eq:interval-conf-mu}
    [\mu(\theta) -\sigma(\theta) q_{\alpha/2}, \mu(\theta) + \sigma(\theta) q_{\alpha/2}]\quad\text{with}\quad\sigma(\theta) = \left(\sum\limits_{k=1}^{K(\theta)}w^{(k)}(\theta)\right)^{-1/2}
\end{equation}
where $\mu$ stands for $\mu_{FE}$~\eqref{eq:overall-es-fe} and $\mu_{RE}$~\eqref{eq:overall-es-re} (depending on the adopted model), and $w^{(k)}$ is either $w^{(k)}_{FE}$~\eqref{eq:weight-fe} or $w^{(k)}_{RE}$~\eqref{eq:weight-re}.
Generally, a vertical line (at $x$-position equal to 0) represents the absence of effect.
If the confidence interval associated with a publication crosses this line, it means that there is no statistically significant difference (in the variable $X$ for the experimental condition $C$) between the two studied groups (e.g\! the study group and the control group) in the considered publication.
Similarly, if the confidence interval of the overall effect size crosses this line, this indicates that the meta-analysis did not find any statistically significant effect between the two studied groups for the considered pool of publications.
In addition to the confidence interval, the z-score $Z(\theta)$ of the overall effect size and the associated p-value are often computed as well:
\begin{equation}
Z(\theta)={\mu(\theta)} / \sigma(\theta)
\end{equation}
where $\mu(\theta)$ can either be $\mu_{FE}$~\eqref{eq:overall-es-fe} and $\mu_{RE}$~\eqref{eq:overall-es-re} and $\sigma(\theta)$ is defined in Equation~\eqref{eq:interval-conf-mu}.
The p-value $p(\theta)$ is given by
\begin{equation}
p(\theta)={2[1-\Phi(|Z(\theta)|)]}
\end{equation}
where $\Phi$ is the standard normal cumulative distribution (two-tailed statistical test).
\subsubsection{Funnel plot}
\begin{figure}[t!]
    \centering
    \begin{minipage}[t]{0.45\linewidth}
    \centerline{\includegraphics[height=0.8\linewidth]{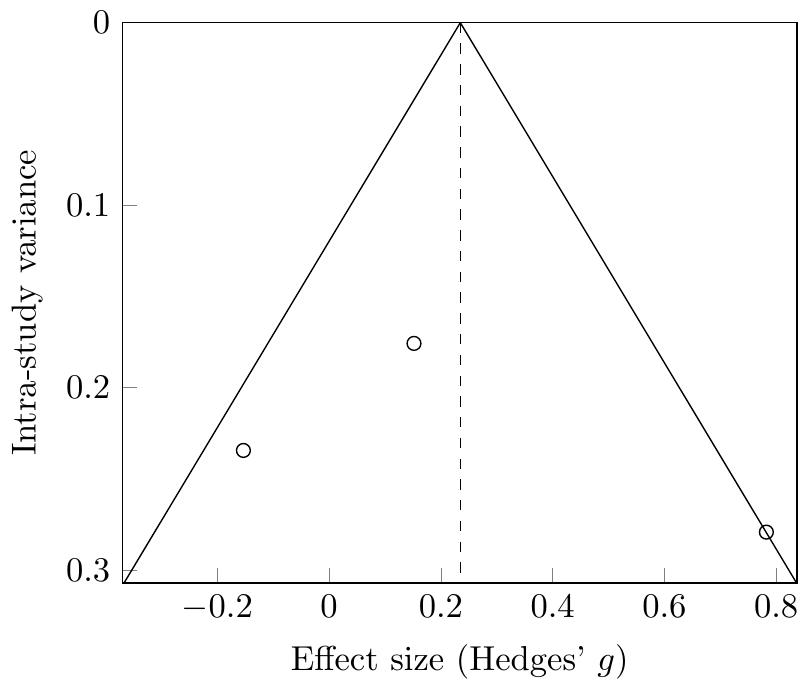}}
    \centerline{\footnotesize (a) Eyes open and retrospective fall recording.}
    \end{minipage}
    \hfil
    \begin{minipage}[t]{0.45\linewidth}
    \centerline{\includegraphics[height=0.8\linewidth, trim=0.5cm 0 0 0, clip]{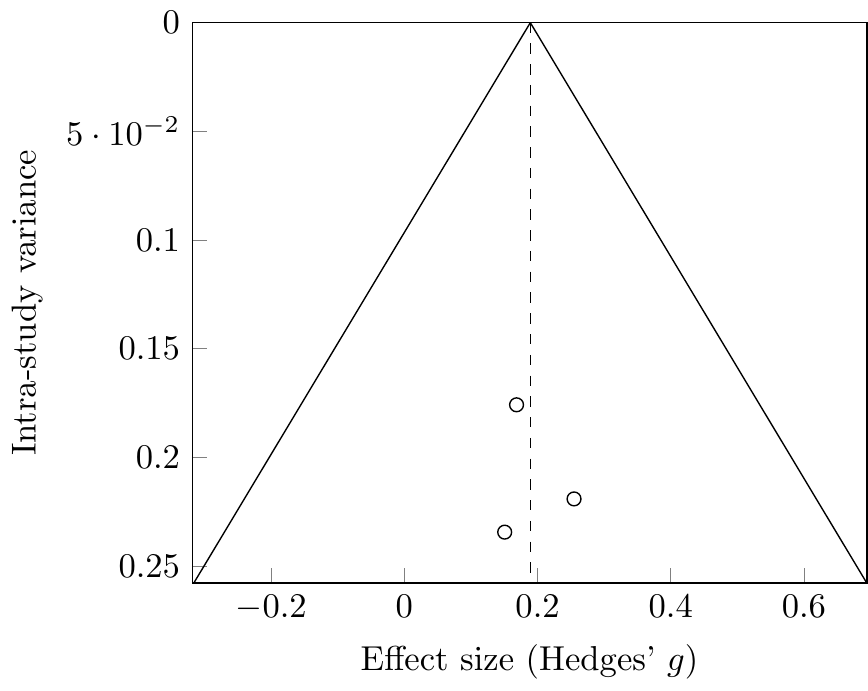}}
    \centerline{\footnotesize (b) Eyes closed and retrospective fall recording.}
    \end{minipage}
    \vskip1em
    \begin{minipage}[t]{0.45\linewidth}
    \centerline{\includegraphics[height=0.8\linewidth]{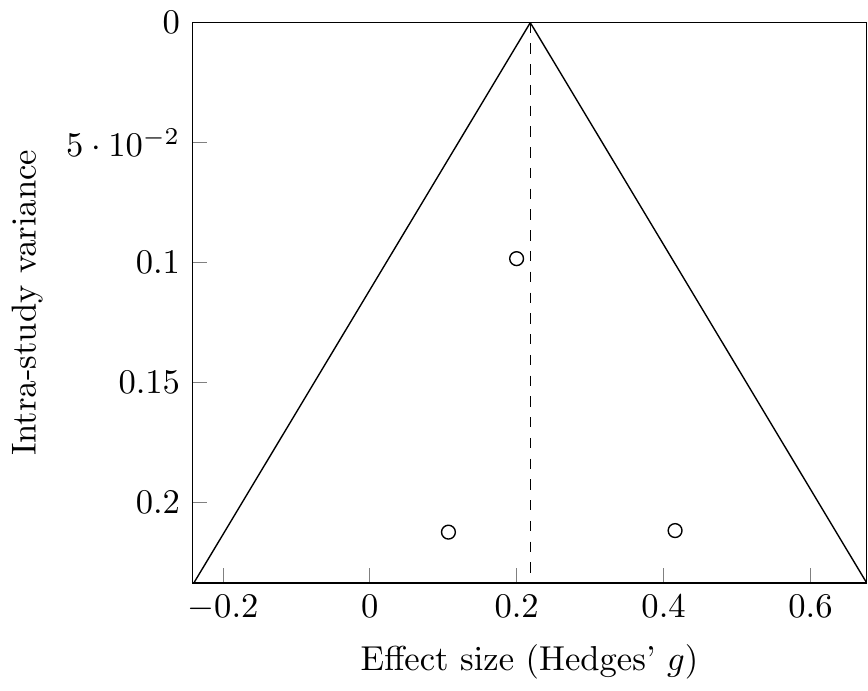}}
    \centerline{\footnotesize (c) Eyes open and prospective fall recording.}
    \end{minipage}
    \hfil
    \begin{minipage}[t]{0.45\linewidth}
    \centerline{\includegraphics[height=0.8\linewidth, trim=0.5cm 0 0 0, clip]{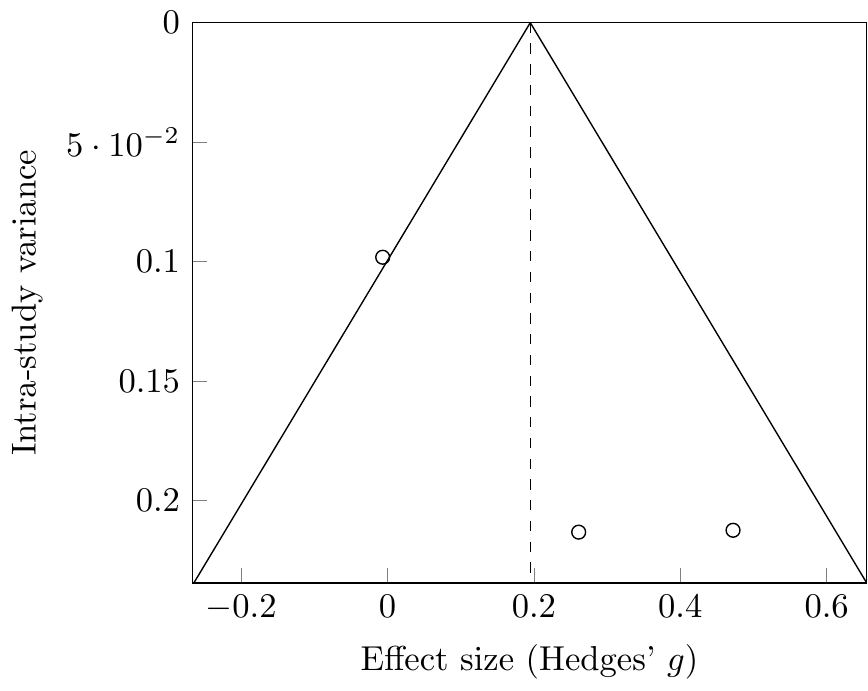}}
    \centerline{\footnotesize (d) Eyes closed and prospective fall recording.}
    \end{minipage}
    \caption{Funnel plot for the input data of Table~\ref{tab:example_data}.
    The variable of interest is `AP mean velocity'.
    Effect size is equal to Hedges' $g$ and the funnel lines are for $\alpha=5\%$.
    The associated forest plot is displayed on Figure~\ref{fig:example-forest-plots}.
    Note that only four out of the eight possible combinations of experimental settings are shown here.}
    \label{fig:example-funnel-plots}
\end{figure}

A funnel plot is a visual tool to assess publication bias.
In a nutshell, publication bias is the consequence of an over-representation of statistically significant results, which can lead to biased effect sizes in meta-analyses~\cite{dubben_systematic_2005,song_publication_2013}.
Formally, a funnel plot is a scatter plot in a two-dimensional plan where the $x$-axis shows the effect size and the $y$-axis, the intra-study variance.
Each publication $k$ ($k=1,\dots,K(\theta)$) with a point of coordinates $(x,y)=(\delta^{(k)}(\theta), \sigma^{(k)}_{\text{intra}}(\theta))$ where $\delta^{(k)}$ can either be $\delta^{(k)}_d$~\eqref{eq:cohen-es} or $\delta^{(k)}_g$~\eqref{eq:hedges-es}, and $\sigma^{(k)}_{\text{intra}}$ is defined in~\eqref{eq:intra-var}.
Usually, the ordinate axis is inverted, so that publications with a large intra-study variance $\sigma^{(k)}_{\text{intra}}$ are below publications with a small intra-study variance.
In addition, the overall effect is graphical represented by two lines forming a funnel, giving its name to this plot, defined by the equations $x = \mu \pm q_{\alpha/2}\times y $ where $\mu$ can be $\mu_{FE}$~\eqref{eq:overall-es-fe} or $\mu_{RE}$~\eqref{eq:overall-es-re} depending on the model (see Section~\ref{sec:fe-or-re}), $\alpha$ is a user-defined level of confidence (usually 5\%), and $q_{\alpha/2}$ is the quantile function of a standard normal distribution.
A vertical line $x=\mu$ is also shown.
\\
Intuitively, publications with a small intra-variability are located in the top of the funnel (the narrow part) while publications with a large variability are dispersed in the bottom of the funnel.
Certain phenomena can easily be seen with a funnel plot.
For instance, over-representation of articles with favourable results would result in a asymmetric distribution of the publications in this representation~\cite{thornton_publication_2000,mlinaric_dealing_2017}. Heterogeneity can also be a source of dispersion and contribute to the asymmetry of the funnel plot~\cite{jin_statistical_2014}.
More interpretations of this representation can be found in~\cite{sterne_funnel_2001,sterne_funnel_2004,lin_graphical_2019}.
An example is shown on Figure~\ref{fig:example-funnel-plots}.

\section{Software description}

\subsection{User input}
User input consists in three elements: the extracted information from the relevant studies (Table~\ref{tab:example_data}), the effect size formula to use (Cohen's $d$ or Hedges' $g$) and the confidence level (usually 5\%).
To work properly, our meta-analysis tool requires a certain formatting of the extracted information, which is now described.
\\
Extracted information are passed using Comma-Separated Values (CSV) files.
Each file contains at least nine columns, separated with semi-colons (``\texttt{;}''), but possibly more, depending on the number of considered experimental conditions:
\begin{itemize}
    \item \texttt{study}: a unique name to identify a study. A common practice is to use ``\{Last name of the first author\}, \{year\}'', for instance, ``Quijoux, 2020''.
    \item \texttt{variable}: a unique name to identify a variable measured during a study.
    \item \texttt{n\_1}: the number $n_1^{(k)}(\theta)$ of participants in the first group.
    \item \texttt{n\_2}: the number $n_2^{(k)}(\theta)$ of participants in the second group.
    \item \texttt{mean\_1}: empirical mean $\mu_1^{(k)}(\theta)$ of the considered variable in the first group.
    \item \texttt{mean\_2}: empirical mean $\mu_2^{(k)}(\theta)$ of the considered variable in the second group.
    \item \texttt{std\_1}: empirical standard deviation $\sigma_1^{(k)}(\theta)$ of the considered variable in the first group.
    \item \texttt{std\_1}: empirical standard deviation $\sigma_2^{(k)}(\theta)$ of the considered variable in the second group.
    \item \texttt{condition\_1}: first condition that defines the experimental setting.
    \item \texttt{condition\_2}, \texttt{condition\_3}, \texttt{condition\_4},\dots: second, third, fourth,\dots, conditions that define the experimental setting. Those columns are optional and should be added only if several experimental conditions are indeed considered. An arbitrary number of columns \texttt{condition\_k} can be added.
\end{itemize}%
Users should be particularly careful to define consistent groups, for instance group 1 can be the experimental/intervention group and group 2 can be the control group.
Under this convention, in the forest plots (see Figure~\ref{fig:example-forest-plots}), the left-hand side (relatively to the vertical dashed line) of the graph favours ``control'' while the right-hand side favours ``experimental/intervention''.
On a more practical note, the columns are separated with semi-colons (``\texttt{;}'') and not commas (``\texttt{,}'').
Common formatting mistakes include: using commas as column separators or decimal delimiters (``1.2'' and not ``1,2''), leaving a semi-colon at the end of a line, using special characters (``\textbackslash'', ``\#'', ``\%'', ``\textdollar'', ``\{\}'', ``\_'', etc.), inconsistent use of upper and lower case (``AP mean velocity'' and ``Ap mean velocity'' and ``ap mean velocity'' are considered as three different variables), leaving an empty string as a condition (it is better to provide a label).
An example input file is displayed on Figure~\ref{fig:raw_input}.
It corresponds to the data of Table~\ref{tab:example_data}.

\begin{figure}[t!]
    \centering\footnotesize
    \begin{verbatim}
                study;variable;n_1;n_2;mean_1;std_1;mean_2;std_2;condition_1;condition_2
                Howcroft, 2015;AP mean velocity;24;76;7.34;2.47;7.65;1.84;EO;Retro
                Howcroft, 2017;AP mean velocity;42;47;7.75;2.15;7.53;1.93;EO;Pro
                Kwok, 2015;AP mean velocity;18;55;1.27;0.45;1.02;0.26;EO;Retro
                Maki, 1994;AP mean velocity;59;37;13;13.7;8.4;3.51;EO;Pro
                Maranesi, 2016;AP mean velocity;63;67;11;6.89;10;6.2;EO;Retro
                Pajala, 2008;AP mean velocity;189;230;8.34;2.81;7.8;2.6;EO;Pro
                Howcroft, 2015;AP mean velocity;24;76;17.34;16.03;15.86;6.74;EC;Retro
                Howcroft, 2017;AP mean velocity;42;47;17.76;13.4;15.11;5.59;EC;Pro
                König, 2014;AP mean velocity;42;42;0.15;1.48;-0.12;0.12;EC;Retro
                Maki, 1994;AP mean velocity;59;37;17.9;15.6;11.9;4.79;EC;Pro
                Maranesi, 2016;AP mean velocity;63;67;16.23;11.27;14.5;9.1;EC;Retro
                Pajala, 2008;AP mean velocity;189;230;12.46;5.09;12.5;6.8;EC;Pro
\end{verbatim}
    \caption{Raw version of Table~\ref{tab:example_data}. This is the input format for the information extracted from the clinical studies of a meta-analysis.}
    \label{fig:raw_input}
\end{figure}

\subsection{Program output}
\begin{figure}[t!]
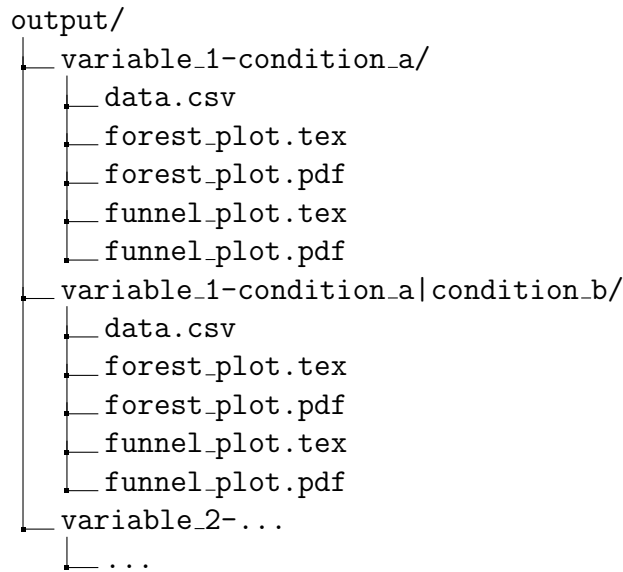

\centering
\begin{minipage}[t]{0.3\linewidth}
\dirtree{%
.1 output/.
.2 variable\_1-condition\_a/.
.3 data.csv.
.3 forest\_plot.tex.
.3 forest\_plot.pdf.
.3 funnel\_plot.tex.
.3 funnel\_plot.pdf.
.2 variable\_1-condition\_a|condition\_b/.
.3 data.csv.
.3 forest\_plot.tex.
.3 forest\_plot.pdf.
.3 funnel\_plot.tex.
.3 funnel\_plot.pdf.
.2 variable\_2-....
.3 ....
}
\end{minipage}
\caption{File structure of a meta-analysis output.}
\label{fig:file_structure}
\end{figure}

The output of our meta-analysis tool is organized in folders, one for each combination of variable and experimental conditions (i.e.\ one per $\theta$).
As an example, for a given variable, e.g.\ ``AP mean velocity'', and a set of experimental conditions, e.g.\ eyes closed (``EC'') and retrospective fall recording (``Retro''), the associated forest and funnel plots can be found in the following folder:
\begin{verbatim}
output/AP mean velocity-EC|Retro/
\end{verbatim}
The name of the variable is separated from the experimental setting by an hyphen (``-'') and the experimental conditions are separated by pipes (``$|$'').
This organization is schematically shown on Figure~\ref{fig:file_structure}.
Within each folder, fives files can be found:
\begin{itemize}
    \item \texttt{data.csv} contains all extracted information (study name, number of participants, empirical means, etc.) and all computed quantities (effect size, confidence interval, etc.) in a tabular form.
    \item \texttt{forest\_plot.pdf} contains the forest plot (see Figure~\ref{fig:example-forest-plots} for examples). In addition, the original \LaTeX code that produced the figure is provided in \texttt{forest\_plot.tex} so that users can tweak the plot to their needs.
    \item Similarly, the funnel plot is given in \texttt{forest\_plot.pdf} and \texttt{forest\_plot.tex} (see Figure~\ref{fig:example-funnel-plots} for examples).
\end{itemize}

\subsection{Interface}

\paragraph{Command-line interface.}
To launch a meta-analysis, the input CSV file (\texttt{input\_file.csv} for instance) should be put in the same folder as the Python files (``\texttt{.py}'' files) and execute the following command in the terminal:
\begin{verbatim}
    python3 main.py --input_fname input_file.csv --alpha 0.05 --which_delta Hedges
\end{verbatim}
The value of $\alpha$ can be changed using the \texttt{--alpha} argument: for $\alpha=0.01$ (i.e.\ 1\%), one only need to replace ``\texttt{--alpha 0.05}'' by ``\texttt{--alpha 0.01}''.
To use Cohen's $d$ instead of Hedges'\ g, again, one only need to substitute ``\texttt{--which\_delta Hedges}'' by ``\texttt{--which\_delta Cohen}''. 

\paragraph{Online demonstration.}

\paragraph{Requirements.} The proposed meta-analysis tool is implemented using well-known open-source languages: Python (\url{python.org}, version 3.6 or more) to compute the effect sizes and confidence intervals, and \LaTeX (\url{tug.org}) to render the forest and funnel plots.
The following Python libraries are needed: pandas, scikit-learn, jinja2, latex, click.
They can easily be installed using \texttt{pip} (\url{docs.python.org/3.6/installing}).

\section{Conclusion}
Meta-analysis can be a useful tool for combining the results of studies dealing with the same issue and having similar methodologies, thus exceeding the individual scope of the selected studies~\cite{russo_how_2007}.
However, the heterogeneity between the included studies is a limitation to the successful conclusion of systematic reviews using this analysis.
It is then necessary to select the studies with the closest experimental conditions.
The aim was to propose a calculation tool in Python programming language due to its wide dissemination.
It is worthy to note that an R package already exists~\cite{del_re_practical_2015} is that although this language is widely disseminated in the scientific community, a library of functions encoded in Python can be useful, as the popularity of this open access language is growing.
The article details the calculus to facilitate the understanding of the process behind the meta-analysis while simplifying the conduct of a sensitivity study.
In the absence of any other library available at the moment, this code provides the rudimentary analysis of a meta-analysis based simply on the data collected through a literature review. 

\section{Acknowledgement}
This study is from the context of a PhD program in partnership with ORPEA group.
This collaboration between Centre Borelli of Paris Descartes University, and ORPEA group is framed in the French conventions for Industrial Training by the Research (CIFRE) managed by the National Association Research-Technology (ANRT).
The authors declare that the research was conducted in the absence of any commercial or financial relationships that could be construed as a potential conflict of interest.
\bibliographystyle{siam}
\bibliography{references}

\end{document}